\documentclass[toc]{PoS}
\usepackage{graphicx}

\title{Precise Numerical Evaluation of the Scalar One-Loop Integrals with 
the Infrared Divergence}

\ShortTitle{Precise Numerical Results of Infrared Divergent Diagrams}

\author{\speaker{F. Yuasa}$^a$, E. de Doncker$^b$, J. Fujimoto$^a$,
N.Hamaguchi$^c$, T. Ishikawa$^a$, Y. Shimizu$^d$\\
\llap{$^a$}High Energy Accelerator Research Organization (KEK), 1-1 OHO Tsukuba, Ibaraki  305-0801, Japan\\
\llap{$^b$}Western Michigan University, Kalamazoo, MI 49008-5371, USA\\
\llap{$^c$}Hitachi, Ltd., Software Division, Totsuka-ku, Yokohama, 244-0801, Japan\\
\llap{$^d$}The Graduate University for Advanced Studies, Sokendai, Shonan Village, Hayama, Kanagawa 240-0193, Japan\\
E-mail: \email{fukuko.yuasa@kek.jp},\\
\email{elise@cs.wmich.edu},\\
\email{junpei.fujimoto@kek.jp}, \\
\email{nobuyuki.hamaguchi.sa@hitachi.com},\\
\email{tadashi.ishikawa@kek.jp},\\
\email{shimiz@suchix.kek.jp}}
       
\abstract
{We present a new approach for obtaining very precise integration 
results for infrared vertex and box diagrams, where the integration 
is carried out directly without performing any analytic integration 
of Feynman parameters. Using an appropriate numerical integration 
routine with an extrapolation method, together with a multi-precision 
library, we have obtained integration results which agree with the 
analytic results to 10 digits even for such a very small photon mass 
as $10^{-150}$ GeV in the infrared vertex diagram.
}

\FullConference{ACAT2007 \\
XI International Workshop on Advanced Computing and Analysis Techniques in Physics Research\\
Amsterdam, The Netherlands\\
April 23-27, 2007}

\begin{document}
\section{Introduction}
In the field of the particle physics more and more precise computation
is required for the progress of the accuracy of the future collider 
experiments. For this reason, the computation of one- and higher loop 
corrections is mandatory. 

General solution has been known for one-loop integrals\cite{tv} and 
there are several excellent programs developed in this decade such as 
{\tt FF}\cite{FF}, {\tt FormCalc-LoopTools}\cite{formcalc}, {\tt XLOOPS-Ginac}\cite{xloop}. 
In parallel, there have been continual efforts to carry out loop integrals by 
fully or semi-fully numerical methods. In early 1990 Fujimoto {\it et al.}\cite{1loop1,supplement,1loop2,1loop3}
have proposed several excellent methods and have shown numerical results. 
In early 2000 Kurihara {\it et al.}\cite{nci} have developed the numerical 
contour method and have shown numerical results. 

Since 2003 we have been developing a method which enables us to carry out the loop integrals in a completely numerical way. In this numerical method, 
to prevent the integral diverging, we put $i \epsilon$ in the denominator 
of the integrand. Here $\epsilon$ is a real positive constant and it is not 
necessary to be infinitesimal. Thanks to this $i \epsilon$, the denominator 
does not vanish and we get the numerical result of $I(\epsilon)$ for 
a given $\epsilon$. Calculating a sequence of $I(\epsilon_{l})$ varying $\epsilon_{l} = \epsilon_{0} * (const.)^{-l}$ ({\it l} = 0,1,2, $\cdot \cdot \cdot$) and extrapolating 
them, we can get the final result of the integration in the limit of $\epsilon \rightarrow 0$.  So far in this method we have calculated several one-loop 
and two-loop integrals and reported the numerical results 
in \cite{dq1, dq2, dq3, dq4}. In these demonstrations
it has been clearly shown that the extrapolation method is very efficient for not only one-loop but also two-loop diagrams.

In the numerical calculation of the loop integrals, it is also very important 
to confirm whether the method can handle the infrared singularities. 
In \cite{dq3}, we have calculated the loop integrals with the infrared 
singularities with a photon mass of up to $10^{-15}$ GeV. In the calculation 
we used the double precision arithmetic for the photon mass of up to $10^{-6}$ 
GeV. However for the photon mass of below $10^{-8}$ GeV the double precision 
is not enough and we use the quadruple precision arithmetic. In the quadruple precision calculation the relative error is given the order of $10^{-7}$.

In this paper, we propose a new approach for our numerical method. 
In this approach we include a precision control technique in addition to the extrapolation method for the loop integral. 
We show that a precision control is mandatory to get the stable results of one-loop diagrams which have infrared singularities with smaller photon mass below $10^{-30}$ GeV. 

The layout of this paper is as follows. In $\S$~\ref{sec:intro} we give formulae of the loop integrals we consider in this paper. In $\S$~\ref{sec:method} we give an explanation of a new approach of our numerical method. Numerical results are shown in $\S$~\ref{sec:results}. Finally in $\S$~\ref{sec:summary} we will summary this paper.

\section{Loop Integrals}\label{sec:intro}
\subsection{One-Loop Vertex}
\begin{figure}
\begin{center}
\includegraphics[width=5cm]{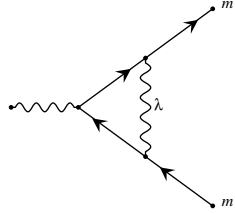}
\caption{One-loop vertex diagram}
\label{fig:vertex}
\end{center}
\end{figure}

For one-loop  vertex diagram in Fig.~\ref{fig:vertex}, the loop integral we consider in this paper is

\begin{equation}
I = \int_{0}^{1} dx \int_{0}^{1-x} dy
{\frac{1}{D}},
\label{eqn:I-vertex}
\end{equation}

where

\begin{equation}
D = -xys + {(x + y)} ^2m^2 + (1 - x - y) \lambda^2.
\label{eqn:D-vertex}
\end{equation}

Here $s$ denotes squared central mass energy and $m$ and $\lambda$
are a mass of external particles and a fictitious photon mass respectively.
We introduce $\lambda$ so as to regularize the infrared singularities.

Replacing $s$ by $s +i \epsilon$ in (~\ref{eqn:D-vertex}), the real part of 
the integral (~\ref{eqn:I-vertex}) becomes 

\begin{equation}
\Re  =
\int_{S}
{\frac{D}{{D}^2 + {\varepsilon}^2 x^{2} y^{2}}}dxdy,
\end{equation}

and the imaginary part becomes

\begin{equation}
\Im  =
\int_{S}
{\frac{\epsilon xy}{{D}^2 + {\varepsilon}^2 x^{2} y^{2}}}dxdy.
\end{equation}

Here $S$ is the triangular region as $0 < x$, $0 < y$ and $x+y < 1$.
Corresponding analytic formulae for one-loop vertex diagrams are 
given in \cite{1loop1,supplement,1loop3}. 

\subsection{One-Loop Box}
\begin{figure}[htb]
\begin{center}
\includegraphics[width=5cm]{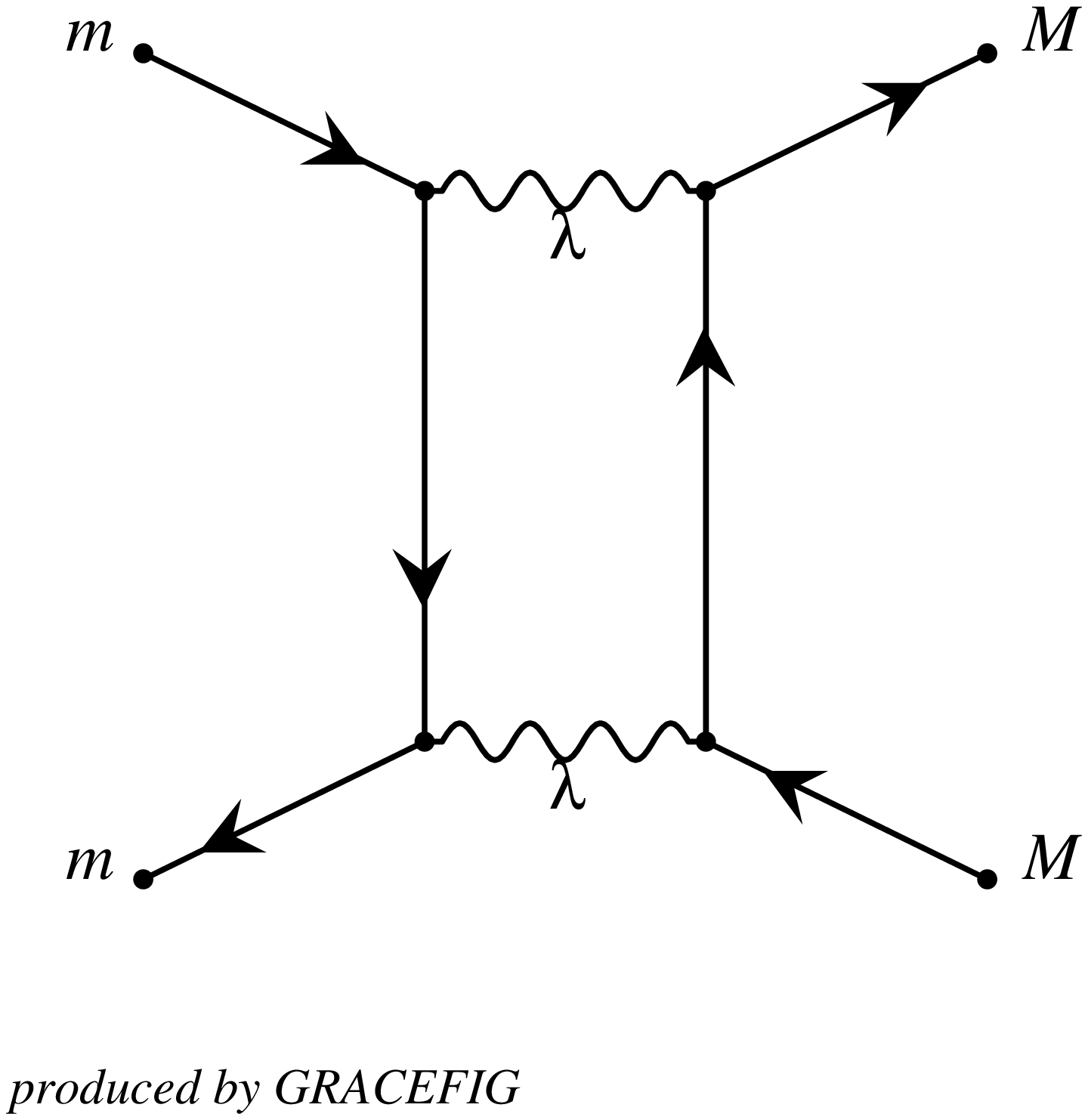}
\caption{One-loop box diagram}
\label{fig:box}
\end{center}
\end{figure}

For one-loop box diagram in Fig.~\ref{fig:box}, the loop integral 
we consider in this paper is 

\begin{equation}
I = \int_{0}^{1} dx \int_{0}^{1-x} dy \int_{0}^{1-x-y} dz
{\frac{1}{D^2}},
\label{eqn:I-box}
\end{equation}

where
\begin{eqnarray}
D = -xys - tz(1-x-y-z)+ (x+y)\lambda^2
+ (1-x-y-z)(1-x-y)m^2 +z(1-x-y)M^2.
\label{eqn:D-box}
\end{eqnarray}

In (~\ref{eqn:D-box})  $s$ denotes squared central mass energy and 
$m$ and $M$ are external masses of particles. 

When we replace $s$ by $s +i \epsilon$ in (~\ref{eqn:I-box}), the real part 
of the integral becomes

\begin{equation}
\Re  = 
\int_{S} 
{\frac{D^2+\epsilon^2 x^2 y^2}{(D^2 + \varepsilon^2 x^2 y^2)^2}}dxdydz,
\end{equation}

and the imaginary part becomes

\begin{equation}
\Im  = 
\int_{S} 
{\frac{2 \epsilon xy D}{(D^2 + \varepsilon^2 x^2 y^2)^2}}dxdydz.
\end{equation}

Here $S$ is the tetrahedron region as $0 < x$, $0 < y$, $0 < z$ and $x+y+z < 1$.
Corresponding analytic formulae for one-loop box diagrams are 
given in \cite{supplement}. 
When we put $M$ = $\mu$, the scalar 
integral for one-loop box diagram becomes

\begin{eqnarray}
\label{eq:1lbgghans}
I_{analytic} &=& - \frac{1}{s} \frac{1}{\sqrt{(-t+\mu^2-m^2)^2-4m^2t}} \nonumber \\
     && \times \ln\left( - \frac{s}{\lambda^2}\right) \ln\frac{(-t+\mu^2+m^2+\sqrt{(-t+\mu^2-m^2)^2-4m^2t})^2}{4m^2\mu^2}.
\end{eqnarray}

%-----------------------------------------------------------------------
\section{Numerical Method}\label{sec:method}
The basic idea of our numerical method is the combination of an efficient
multi-dimensional integration routine and an extrapolation method.
Using this method we have shown several numerical results of loop integrals 
and that they show good agreements with analytical results.
In the calculation so far we have used double precision arithmetic or quadruple precision arithmetic if necessary for one-loop and two-loop integrals.
However for the infrared divergent diagram even with the quadruple precision arithmetic it becomes harder to get a result with an enough accuracy as the value of the $\lambda$ becomes much smaller such as $10^{-15}$ GeV\cite{dq3}.
Therefore in our new approach we include the concept of a precision control.

 In the following subsections, we give a very simple explanation of the multi-dimensional integration routine, an extrapolation method and a multi-precision library we used in our  
new approach respectively.

\subsection{Multi-dimensional Integration Routine}\label{subsec:integ}
We use {\tt DQAGE} routine in the multi-dimensional integration.
It is included in the package {\tt QUADPACK} \cite{QUADPACK} and it is a globally adaptive integration routine. Though it is a routine for one-dimensional integration, 
we use it iteratedly \cite{iterate}. 

\subsection{Extrapolation Method}\label{subsec:extrapolation}
Although there are several ways to accelerate the conversion of the sequences.
Of them, the $\epsilon$ algorithm by P.Wynn\cite{wynn} is recommended as 
the best all-purpose method for slowly converging sequences.
It is an implementation of Shanks' transformation \cite{shanks} and is
represented in a recursive formulae. 

\subsection{Multi-Precision Library}\label{subsec:precision}
When we know correctly whether the precision arithmetic used in the calculation is sufficient or not, we can select the necessary precision arithmetic and the results become reliable.
J.Fujimoto {\it et. al} have discussed the importance of a precision control in \cite{acat05JF}. We use {\tt HMlib}\cite{hmlib} as the multi-precision library in our new approach because it gives an information of the lost-bits during the calculation and we can guarantee the precision of the results. 
Taking the calculation of one-loop vertex diagram as an example, we show the information of the lost-bits supplied by {\tt HMlib} in Table~\ref{tab:hmlib}.

\begin{table}
\caption{In {\it P}-precision presentation implemented in {\tt HMlib}, a sign bit is 1 bit and an exponent bit is 15 bits and a mantissa is ($32 \times P - 16$) bits. When $P = 4$ (quadruple precision), the mantissa becomes 112 bits. In the Table when $\lambda$ is $10^{-23}$ GeV, the maximum number of lost-bits is the same as the number of bits of the mantissa.}
\label{tab:hmlib}
\begin{center}
\begin{tabular}{|l|r|r|} \hline
$\lambda$ [GeV] & Average lost-bits & Maximum lost-bits\\ \hline 
$10^{-20}$ & 88 & 92\\ \hline
$10^{-21}$ & 98 & 102\\ \hline
$10^{-22}$ & 108 & 112\\ \hline
\end{tabular}
\end{center}
\end{table}

\section{Numerical Results}\label{sec:results}
\subsection{One-Loop Vertex}

The numerical results are shown in Table~\ref{tab:E-PT}. 
Results are compared to analytic results evaluated by the formulae in \cite{1loop3}.
As an example, in Table ~\ref{tab:DQ-EL150} we show the parameters we used in the calculation which are {\tt key} and {\tt limit} used in {\tt DQAGE}, starting $\epsilon$ and ending $\epsilon$ for an extrapolation method.

%-----------------------------------------------------------------------
\begin{table}
\caption{Numerical results of one-loop vertex diagram with $\sqrt{s} = 500$ GeV, $t = -150^{2}$ $GeV^{2}$, $m = m_{e} = 0.5 \times 10^{-3}$ GeV. 
The upper is the result of Real part and the lower is one of Imaginary part.}
\label{tab:E-PT}
\scriptsize{
\begin{center}
\begin{tabular}{|l|r|r|r|r|} \hline
$\lambda$ [GeV]& Numerical Results & Precision & Analytical Results [quadruple precision]& Agreement\\ \hline
$10^{-30}$ & -0.150899286980769753D-01 $\pm$ 0.771D-26&8& -0.15089928698048229151911707856807798E-01 & 12\\
$        $ &   0.189229839615898822D-02 $\pm$ 0.124D-25&8& 0.18922983961552538934533667780832051E-02 & 12\\ \hline
$10^{-60}$ & -0.303593952562854951D-01 $\pm$ 0.178D-16&16& -0.30359395256226015422329307055958645E-01 & 12\\
$        $ &   0.362840665514227622D-02 $\pm$ 0.896D-13&16&  0.36284066551349654484509779501460548E-02 & 11\\ \hline
$10^{-80}$ & -0.405390396284235075D-01 $\pm$ 0.580D-15&16& -0.40539039628344539602607706522058889E-01 & 11\\
$        $ &   0.478581216125478532D-02 $\pm$ 0.401D-12&16&  0.47858121611214398184493853981879066E-02 & 10\\ \hline
$10^{-120}$& -0.608983283726997427D-01 $\pm$ 0.556D-15&32& -0.60898328372581587963164505454259761E-01 & 11\\
$         $&   0.710062317325786663D-02 $\pm$ 0.415D-12&32&  0.71006231730943885584462002942716100E-02 & 10\\ \hline
$10^{-150}$& -0.761677949309069452D-01 $\pm$ 0.931D-15&32& -0.76167794930759374233582104516222189E-01 & 11\\
$         $&   0.883673143185801414D-02 $\pm$ 0.260D-13&32&  0.88367314320741001134438114507364254E-02 & 10\\ \hline
$10^{-160}$& -0.812576170752810666D-01 $\pm$ 0.549D-10&32& -0.81257618347269926993757808016127325E-01 &7 \\
$         $&   0.941543418501223556D-02 $\pm$ 0.109D-11&32&  0.94154343249672244271406905030553399E-02 & 7\\ \hline
\end{tabular}
\end{center}
}
\end{table}
\normalsize
%-----------------------------------------------------------------------
\begin{table}
\caption{Parameters used in {\tt DQAGE} and in an extrapolation method for the one-loop vertex diagram with $\sqrt{s} = 500$ GeV, $m = m_e = 0.5 \times 10^{-3}$ 
GeV, $\lambda = 10^{-150}$ GeV. The upper is the parameters for the Real part and the lower is ones for Imaginary part.}
\label{tab:DQ-EL150}
\begin{center}
\begin{tabular}{|r|r|r|r|} \hline
$key_{x}$ & $limit_{x}$ & $\epsilon_{start}$ & $\epsilon_{end}$\\
$key_{y}$ & $limit_{y}$ &                    &                 \\ \hline
1         & 600         & 0.36572620D+04     &0.28485158D+03\\
1         & 600         &                    &\\ \hline
1         & 600         & 0.36572620D+04     &0.28485158D+03\\
1         & 600         &                    &\\ \hline
\end{tabular}
\end{center}
\end{table}
%-----------------------------------------------------------------------

\subsection{One-Loop Box}
The numerical results of one-loop box diagram shown are shown in Table~\ref{tab:gg-PT}.
The results are compared to the analytic results evaluated by (~\ref{eq:1lbgghans}) in a quadruple precision. 
%When $\lambda$ is $10^{-5}$ in Table~\ref{tab:gg-PT}, 
%the agreement between numerical results and analytic results 
%is not good enough compared to ones
%with much smaller $\lambda$ such as $10^{-9}$. 
%This is because $\lambda$ is not enough small to neglect 
%$\lambda^2O(1/\lambda \ln{\lambda})$ in the analytic formulae 
%when $\lambda$ is $10^{-5}$.
As an example, in Table ~\ref{tab:DQ-gg25} we show the parameters we used in the calculation which are {\tt key} and {\tt limit} used in {\tt DQAGE}, starting $\epsilon$ and ending $\epsilon$ for an extrapolation method.

%-----------------------------------------------------------------------
\begin{table}
\caption{Numerical results of the loop integral of one-loop box diagram with $\sqrt{s} = 500$ GeV, $t$ = -$150^2 GeV^2$, $m$ = 0.5 $\times 10^{-5}$  GeV, $M$ = 150 GeV. This is results of Real part.}
\label{tab:gg-PT}
\scriptsize{
\begin{center}
\begin{tabular}{|l|r|r|r|r|} \hline 
$\lambda$ [GeV]& Numerical Results & Precision & Numerical Results [quadruple precision]& Agreement\\ \hline
$10^{-15}$& -0.1927861102670278D-06 $\pm$ 0.314D-14 &4,4,2& -0.19278611224396411606895771777195708E-06 & 8\\ \hline
$10^{-20}$& -0.2472486348234972D-06 $\pm$ 0.586D-15 &4,4,2& -0.24724863525991758999379683385554222E-06 & 9\\ \hline
$10^{-25}$& -0.3017111253761463D-06 $\pm$ 0.111D-13 &4,4,2& -0.30171115827587106391863594993913029E-06 & 7\\ \hline
$10^{-30}$& -0.3562028882831722D-06 $\pm$ 0.867D-10 &4,4,2& -0.35617368129182453784347506602271837E-06 & 4\\ \hline
\end{tabular}
\end{center}
}
\end{table}
\normalsize
%-----------------------------------------------------------------------
\normalsize
\begin{table}
\caption{Parameters used in {\tt DQAGE} and in an extrapolation method for the one-loop box diagram with  $\sqrt{s} = 500$ GeV, $t$ = -$150^2 GeV^2$, $m = 0.5 \times 10^{-3}$ GeV, $M$ = 150 GeV and $\lambda = 10^{-25}$ GeV.} 
\label{tab:DQ-gg25}
\begin{center}
\begin{tabular}{|r|r|r|r|} \hline 
$key_{x}$ & $limit_{x}$ &$\epsilon_{start}$& $\epsilon_{end}$\\
$key_{y}$ & $limit_{y}$ &                  &                 \\
$key_{z}$ & $limit_{z}$ &                  &                 \\ \hline
1         & 400         & 0.36572620D+04   & 0.28485158D+03  \\ 
1         & 400         &                  &                 \\
1         & 100         &                  &                 \\ \hline
\end{tabular}
\end{center}
\end{table}
\normalsize
%-----------------------------------------------------------------------

\subsection{Elapsed Time}
In Table~\ref{tab:elapsevertex} and ~\ref{tab:elapsebox}, the elapsed times required in the several typical calculations are shown.
\begin{table}
\caption{Elapsed time required in the calculation of the one-loop vertex diagram}
\label{tab:elapsevertex}
\begin{center}
\begin{tabular}{|l|r|r|l|l|}\hline
$\lambda$ [GeV] &Real Part& Imaginary Part&Precision& CPU \\ \hline
$10^{-30}$&1.8days &1.8days &8,8 & Xeon 3.06GHz \\ \hline
$10^{-150}$&6.8days &6.7days &32,32 &Xeon 3.06GHz \\ \hline
\end{tabular}
\end{center}
\end{table}

\begin{table}
\caption{Elapsed time required in the calculation of the one-loop box diagram}
\label{tab:elapsebox}
\begin{center}
\begin{tabular}{|l|r|l|l|}\hline
$\lambda$ [GeV] &Real Part& Precision& CPU \\ \hline
$10^{-15}$&1.5days &4,4,2 &Opteron 2.2GHz \\ \hline
$10^{-25}$&3.0days &4,4,2 &Opteron 2.2GHz \\ \hline
\end{tabular}
\end{center}
\end{table}

\section{Summary}\label{sec:summary}
We have shown that a new numerical approach presented in this paper is applicable to
the scalar one-loop vertex and box diagram with infrared singularities.
The new approach consists of an integration routine, an extrapolation 
method and a precision control. Several numerical results are shown and they agree with analytic ones. All the calculation of the one-loop vertex diagram are carried out in at least
quadruple precision arithmetic and some are in 8-, 16- and 32-precision arithmetic. 
From these demonstrations, it is confirmed that a precision control plays an important role in the calculation of the loop integrals with the infrared singularities. 
Although higher precision arithmetic costs a lot of CPU time, 
it supplies the information of the lost-bits occurred during the calculation and this is extremely valuable in getting the reliable results. 
To reduce CPU time we will be able to use the parallel computing technique.

\acknowledgments
We wish to thank the members of MINAMI-TATEYA collaboration for discussions.
We also wish to thank Prof. Kaneko and Dr. Kurihara
for their valuable suggestions. We wish to thank Prof. Kawabata for his encouragement and support. This work was supported in part by the Grants-in-Aid (No.17340085 and No.17540283 ) of JSPS.

\end{document}